\documentclass[preprint]{aastex}

\newif\ifAMStwofonts
\AMStwofontstrue
 
\def\kms{km~s$^{-1}$}

\def\degree{$^{\circ}$}

\def\ga{\mathrel{\hbox{\rlap{\hbox{\lower4pt\hbox{$\sim$}}}\hbox{$>$}}}}
\def\la{\mathrel{\hbox{\rlap{\hbox{\lower4pt\hbox{$\sim$}}}\hbox{$<$}}}}
 
\shorttitle{}
 
\shortauthors{S.\ Stanimirovi\'{c} et al.}
\slugcomment{{\it Accepted for publication in ApJ}} 
\begin{document}
 
\title{The small-scale structure of the Magellanic Stream}
 
\author{Sne\v{z}ana Stanimirovi\'{c}}
\affil{Arecibo Observatory, NAIC/Cornell University, HC 3 Box 53995,
Arecibo, PR 00612}
\email{sstanimi@naic.edu}
\author{John M. Dickey}
\affil{Department of Astronomy, University of Minnesota, 116 Church St. SE,
Minneapolis, MN 55455}
\author{Marko Kr\v{c}o}
\affil{Department of Physics and Astronomy, Colgate University, 
13 Oak Drive,  Hamilton, NY 13346}
\author{Alyson M. Brooks}
\affil{Columbia Astrophysics Laboratory, Columbia University, 
550 West 120th Street, New York, NY 10027}

\begin{abstract}

We have mapped two regions at the northern tip of
the Magellanic Stream in neutral hydrogen 21-cm emission 
using the Arecibo telescope. The new data are
used to study the morphology and properties of the Stream far away from the
Magellanic Clouds, as well as to provide indirect constraints on the
properties of the Galactic Halo. We investigate confinement
mechanisms for the Stream clouds and conclude that these clouds cannot 
be gravitationally confined or in free expansion. 
The most likely mechanism for cloud confinement is
pressure support from the hot Galactic Halo gas. This allows us to place an
upper limit on the Halo density: $n_{\rm h}(15~{\rm kpc}) = 10^{-3}$ 
cm$^{-3}$ and/or $n_{\rm h}(45~{\rm kpc}) = 3 \times 10^{-4}$ cm$^{-3}$
depending on the distance. These
values are significantly higher than predicted for an isothermal
stratified Halo. 
\end{abstract}
 
\section{Introduction}

The Magellanic Stream (MS)
is a thin, $\sim$ 10\degree~wide \citep{Putman99}, tail of neutral
hydrogen (HI), emanating from the Magellanic Clouds (MCs: the Large Magellanic
Cloud, LMC, and  the Small Magellanic Cloud, SMC) and trailing away for
almost  100\degree~across the sky ($-60$\degree $<$ Dec $<+15$\degree).   
This huge
HI structure is the most  fascinating signature of the  wild past
interaction of our Galaxy with the MCs, and the MCs with each other. 
At the same time, streaming so far away from the
MCs into the Galactic Halo, the MS is an excellent probe of the
properties of the outer Galactic Halo, for which otherwise we have very few
observational clues.

For many years the MS was viewed as a complex of six  discrete
concentrations \citep{Mathewson74}.  The new HI Parkes All-Sky Survey
(Putman \& Gibson 1999) reveals the more complex  nature of the MS,  with a
fascinating network of  filaments and clumps. No stars have been found so
far in the MS, making it the closest medium to a primordial environment that
exists in the Local Group today. 

H$_{\alpha}$ emission appears to be
routinely detected throughout the MS \citep{Weiner02}, suggesting that
the MS is being ionized by photons escaping from the Galaxy
\citep{Bland-Hawthorn99}.  While this mechanism is in accord with the
H$\alpha$ observations of many high velocity clouds (HVCs), the predicted
H$\alpha$ flux for the MS is significantly lower than what is 
found observationally \citep{Weiner02,Bland-Hawthorn02}, 
suggesting that an additional source of ionization must be invoked.
Several possible mechanisms have been suggested, including  an interaction
between the MS gas and the hot Galactic Halo gas,
turbulent mixing, the existence of young massive stars embedded in 
the MS \citep{Hawthorn-Putman01}, shocks, self-interaction of the MS gas, 
and magnetic fields \citep{Konz01}.

The MS appears to be
the result of interaction between the Galaxy and the MCs, but there is no
consensus as to the exact
form of this interaction \citep{Putman00}.
Theories have swung back and forth on the relative  importance of tidal
stripping \citep{Murai80, Gardiner96}  and various
kinds of gas dynamical interactions \citep{Mathewson87}. The recent
discovery by \cite{Mary-nature} of a counter HI stream,  leading the MCs, 
lends weight to the tidal origin theory.  \\
 
The work presented in this paper was motivated by the following 
two questions.

(1) {\it Is the MS dissipating into the Halo?}\\ 
The most fundamental issue about the origin and structure  of the 
MS is to what extent interaction
with the Galactic Halo determines or influences the MS gas.  This problem
becomes particularly important at the extreme northern end of the MS,
because without pressure from an external medium the MS clouds should
dissipate on a short time scale.  However, if the external pressure is 
sufficient to confine the MS clouds, then the gaseous Galactic Halo has a much
higher scale height than most theories predict. 
This  paper addresses several issues in relation to this particular question.

(2) {\it Is a hierarchy of structures present in this almost primordial
environment?}\\ 
In the Galaxy, a hierarchy of structure is present in the diffuse
interstellar medium (ISM), down to very small
scales. This is mainly ascribed to interstellar (IS) turbulence.  This
process is seen both in the ionized medium, traced by pulsar scintillation
and dispersion variations, and in the atomic gas, traced by the spatial
power spectrum of HI emission and absorption fluctuations 
\citep{Fiedler94,Green93,Lazarian99,Dickey01}.  Recently it has become 
possible to extend
these studies to other galaxies, at least to our neighbors 
\citep{Westpfahl99,Stanimirovic99,Elmegreen00a}. 
The results suggest that the processes we see at work in the solar neighborhood
are quite general.  As the IS turbulence is driven, at least partially, 
by stellar processes such as
winds and supernova remnants, it would be quite a surprise if the same
turbulence spectrum was seen in an environment without any stars at
all, like the MS. We have addressed this question 
briefly in \cite{Stanimirovic02-IAP} and will
concentrate on it  further in a subsequent paper.

This paper is organized as follows:
In Section~\ref{s:previous} we summarize previous HI observations of the
MS, and especially its northern part. As parameters such as the MS age 
and its distance come up often throughout this paper, a brief 
description of various theories for the MS formation is also outlined here.
Section~\ref{s:observations} describes new HI observations of two regions
at the tip of the MS conducted with the Arecibo telescope.
The HI data and observational results obtained are presented in 
Section~\ref{s:observational-results}.  Several clumps and
their properties are shown in
Section~\ref{s:clump-identification+analyses}.
In Section~\ref{s:discussion} we discuss clump confinement issues 
and the constraints they place on the
density of the Galactic Halo. Expected Halo
densities, as well as several theoretical approaches for considering 
interactions between an MS cloud and the ambient medium, are discussed. 
An alternative explanation for the velocity field of one of the 
regions is also presented. We summarize our results 
in Section~\ref{s:conclusions}.

\section{Previous HI observations and theoretical models}
\label{s:previous}

\begin{figure*}
\epsscale{1.0}
\caption{\label{f:mary_for_snez}
{\bf [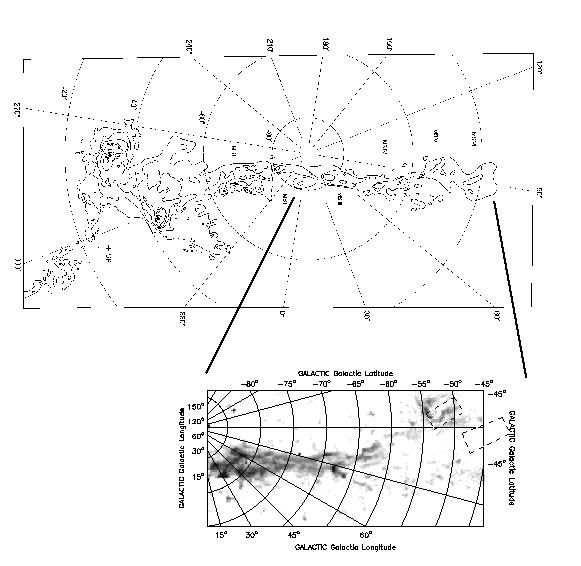]} The HI distribution of the MS from Mathewson \& 
Ford (1984) is shown in the
upper panel. The lower panel shows clouds
MS III through MS VI of the MS as seen by the Parkes HIPASS
survey (Putman et al. 2002).
The two regions observed with the Arecibo telescope and presented in this paper
are outlined with dashed lines.}
\end{figure*}
 
\subsection{Lower resolution observations}
First HI detections of the MS were obtained by \cite{Dieter65}
and \cite{Wannier72}, however \cite{Mathewson74} were the first to discover a
coherent large-scale structure which they named the Magellanic Stream.  
Observations with better 
resolution and sensitivity followed by \cite{Cohen75},
\cite{Mirabel79}, \cite{Erkes80}, \cite{Mirabel81} and
\cite{MathewsonFord84}.  These observations revealed the large-scale
morphology of the MS, that is traditionally viewed as a long filament,
containing a bead-like sequence of six discrete clouds: MS I, near the MCs 
at Dec $\sim -60$\degree, to MS VI at the very northern tip at Dec 
$\sim +15$\degree~\citep{Mathewson77}, see the top panel in Fig. 1. 
Mirabel (1981) in particular focused on the tip of the MS, searching
for evidence of its disintegration. \cite{Wayte89} used the 
Parkes 64-m dish to uncover complex and
chaotic motions right at the MS tip.

The most recent HI observations of the MS, with spatial resolution of
15 arcmin and velocity resolution of 26 \kms, were undertaken with the
Parkes Multibeam system \citep{Putman00,Putman02}.  These, as 
well as the recent
observations obtained with the same instrument but with higher velocity
resolution of 1 \kms~by \cite{Bruns00}, revolutionized the classical
picture of the MS.  Instead of six concentrations of gas, two striking, long
and distinct filaments are seen  that run together for most of the length
of the MS but merge in several places, giving the appearance of a
double-helix structure. North of Dec $\sim 0$\degree, the two filaments break
into a network of clumps and smaller filaments (see the bottom panel in 
Fig.~\ref{f:mary_for_snez}).
A global velocity gradient is observed along the MS, from $\sim +350$
\kms~(LSR) near the MC's to $\sim -450$ \kms~(LSR) at the very northern tip.
 
\subsection{Theoretical models}
\label{s:theories}
  
The two leading hypotheses for explaining the origin of the
MS are based on tidal interaction versus ram pressure stripping, 
and are applied in 
several models. These two families of models predict
different ages and distances for the MS. As the predicted 
parameters are important for
the calculations in Sections~\ref{s:clump-identification+analyses} 
and~\ref{s:discussion} we carry out calculations for 
both sets of model predictions. 
  
{\bf The tidal hypothesis} 
\citep{Murai80,Lin82,Gardiner94,Gardiner96,Yoshizawa99} ---
invokes the idea of gravitational stripping of the SMC gas by the Galaxy.
According to N-body simulations, the MS was formed 1.5 Gyr ago in a tidal
encounter between the LMC and the SMC at their perigalactic passage. 
This encounter drew gas out of  the SMC that later evolved into a leading
bridge and a trailing tail.  The MS follows
the MCs on an almost polar orbit with a transverse velocity  of 
220 \kms~(LSR). Simulations show that the MS 
gas occupies a range of distances, from 45 kpc in the vicinity of 
MS III, to 60 -- 70 kpc at MS VI.
  
One of the main drawbacks of the tidal model used to be
the observation of only one tail of gas behind the MCs. Recent Parkes
observations \citep{Mary-nature} have discovered the `Leading Arm Feature'
(LAF), a counter-stream that leads the direction of motion of the MCs,
as expected in
the tidal models. The absence of stars in the MS
is another troubling issue for tidal models, as both stars and gas should 
be affected equally by tides.  However, the
most recent N-body simulations by \citep{Yoshizawa99} show that a very compact
initial configuration of the SMC stellar disk could result in only gas
being disrupted, while the stars are left unaffected.

{\bf The ram-pressure hypothesis} \citep{Moore94,Heller94,Sofue94} --- 
invokes the idea of ram-pressure stripping of gas
from the MCs by an extended halo of diffuse gas around the Galaxy.  
Here, the MCs entered the extended ionized halo of the Galaxy 
some 500 Myr ago at a galactocentric distance of 65 kpc. 
Material was then stripped from the SMC
and the intercloud region and began to fall toward the Galaxy.  In this
model, the gas with the lowest column density, that lost the most 
orbital angular momentum, has fallen the furthest toward the Galaxy. 
This gas is currently located at the tip of the MS, at a 
galactocentric radius of 25 kpc. The model requires that
the diffuse Galactic Halo has densities of $\sim 5\times10^{-5}$ 
cm$^{-3}$ at distance of 65 kpc in order to strip gas, 
but not decelerate it too much.

{\bf The tidal $+$ drag model} (Gardiner 1999; Moore \& Davis, in
preparation) The recent observational confirmation of the LAF provides a
new set of parameters for theoretical models. In order to better constrain
the observational characteristics of the LAF, \cite{Gardiner99}  introduced
a weak non-gravitational term (a drag term) into the classical tidal
model. The new simulations show interesting  results: while the LAF is
reproduced,  simulated particles form a complete ring from the LAF,
connecting the LAF to two trailing streams.  These streams have different
galactocentric velocities starting from MS I to MS V, and then meet
together again around MS VI (see Fig. 4 from Gardiner 1999).  This
simulation predicts that large-scale velocity splitting should be seen
throughout most of the MS.  The success of this model, based on both
gravitational and non-gravitational forces, encourages further developments
in the field that are eagerly anticipated (Moore,  in preparation;
Maddison, Kawata, \& Gibson 2002).

\section{Observations and data reduction}
\label{s:observations}
 
The observations were obtained with the 305-m
Arecibo telescope\footnote{The Arecibo Observatory is part of the 
National Astronomy and Ionosphere Center, which is operated by Cornell 
University under a cooperative agreement with the National Science 
Foundation.} in June and July of 2000 and June of 2001. The
Gregorian feed was used with the narrow 21-cm (L-band) receiver. The
illuminated part of the 305-m dish covers an area of about 
210$\times$240 m, giving a beam FWHM of approximately 3$'$.1 $\times$
3$'$.4 arcmin
\citep{Heiles01pasp}. Two observing bandwidths of 12.5 and 6.25 MHz were
used simultaneously on two correlator boards, each with 1024 channels,
for each circular polarization. The resultant velocity resolution
is 2.6 and 1.3 \kms, respectively.
  
Data were taken in the on-the-fly mapping mode, by driving in RA, at a rate of
4 arcsec per sec, and stepping in Dec by 2 arcmin between
strips. The correlator was set
to make one scan per strip by recording data continuously every 30 sec,
which is effectively every 2 arcmin on the sky.  As a result the data were
slightly undersampled.  At the beginning of each scan, a noise
diode of known temperature was fired for 3 sec for calibration.

Two regions of interest were selected from Mirabel et al. (1979): 
one centered at
the northern tip of the MS, $(l,b)=(86.7,-42.8)$, and the other in the MS V
region, $(l,b)=(94.4,-50.6)$ (Fig. 1, these regions are outlined in the
lower panel).  Below we refer to
these areas  as MS VI and MS V, with the understanding that our maps do not
cover the full extents of those clouds.  Both regions were   mosaiced with
many small, overlapping maps, each of 10$\times$10 arcmin, which take $\sim$  1
h to observe. The mosaic pieces were combined during the gridding process.

We removed a linear baseline by fitting ranges of
emission-free channels on both sides of the line. 
The gain correction, as a function of both
azimuth and zenith angle, was then applied. All spectra were
subsequently convolved onto a sky grid using a Gaussian convolution
function with  FWHM of 2.6 arcmin, broadening the angular resolution
to $\sim$ 4 arcmin.  A gridding correction factor of 1.45
\citep{SramekSchwab} was
applied to account for the new beam size (S. Stanimirovic 2002, 
in preparation). 
  
The relationship between flux density and brightness temperature is given
by $S({\rm Jy~beam^{-1}})=0.11T_{\rm b} (\rm{K})$.  The noise level is 0.04 K
per 1.3 \kms~wide channel, which is equivalent to a column density of 
$\sim$ 10$^{17}$ cm$^{-2}$. The peak brightness temperature is 1.5 K
in the MS VI region, and 1.1 K in MS V.  For comparison,
Mirabel et al. (1979) obtained a maximum column
density of $\sim$ 4$\times$10$^{19}$ cm$^{-2}$ in the MS VI region 
using the Jodrell Bank telescope. In the same region we derive a 
maximum column density of 4.3$\times$10$^{19}$ cm$^{-2}$.

\section{Observational Results}
\label{s:observational-results}

\subsection{Morphology and velocity field of the MS VI region}

\begin{figure*}
\epsscale{0.9}
\caption{\label{f:ms1_channels}
{\bf [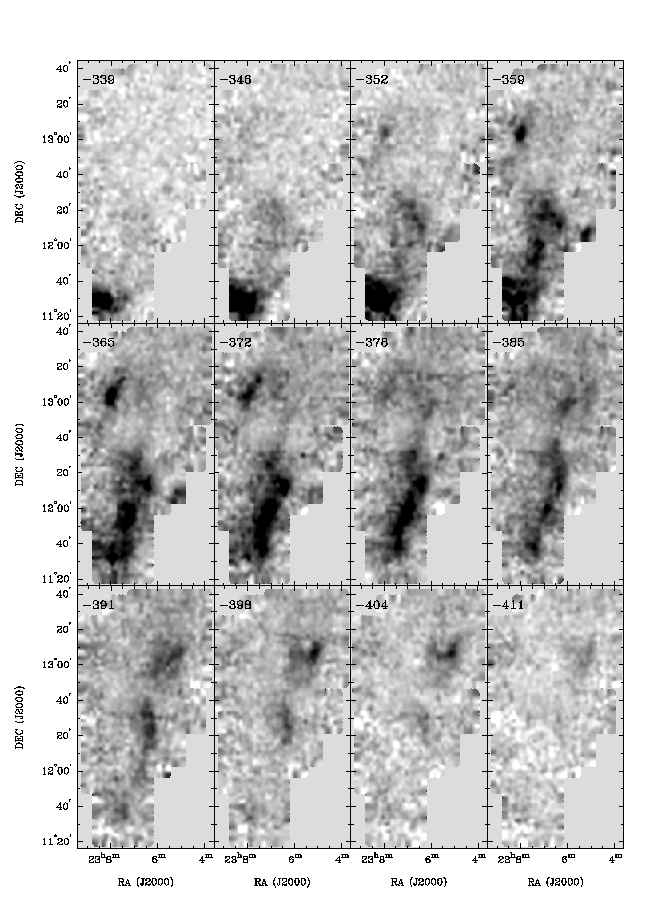]}The RA--Dec images of the MS VI region for
different LSR velocities given in the top-left corner. The grey-scale
intensity range is $-0.05$ to 0.3 K with a linear transfer function. }
\end{figure*}

\begin{figure*}
\epsscale{0.95}
\caption{\label{f:ms1_posvel} 
{\bf [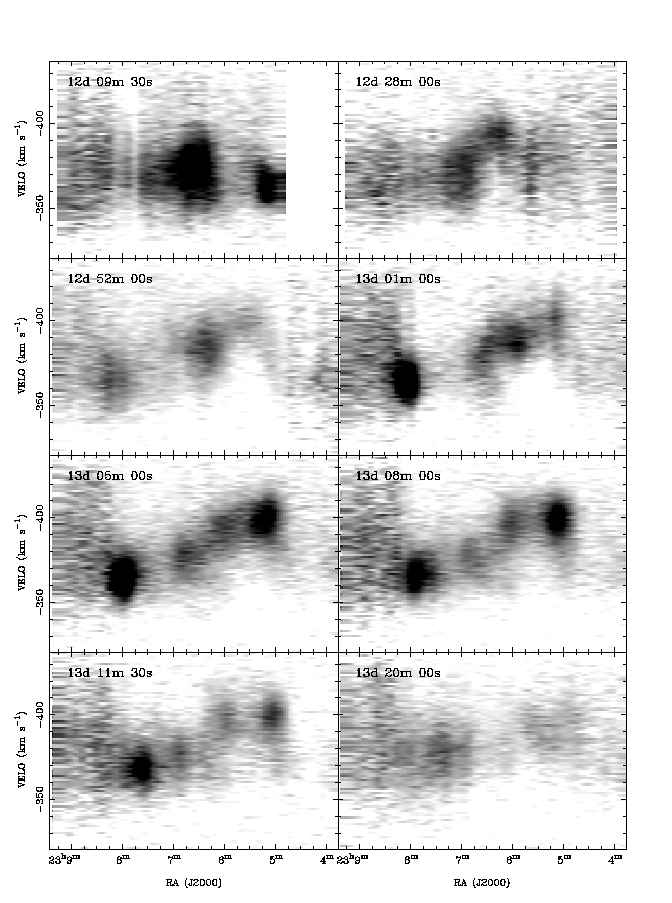]} Position--velocity diagrams of the MS VI 
region taken at different Dec values,
given in the top-left corner. The grey-scale intensity range is 0.05 to 0.7
K with a linear transfer function. }
\end{figure*}

\begin{figure*}
\epsscale{1.0}
\caption{\label{f:ms1_0mom} 
{\bf [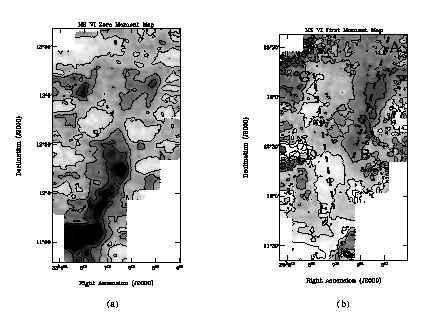]}(a) An HI column density image of MS VI overlaid with contours.
The grey-scale range is 0 to 2.4$\times10^{19}$ cm$^{-2}$.
Contour levels are from 7 to 30$\times10^{18}$ cm$^{-2}$, with a
contour step of 6$\times10^{18}$ cm$^{-2}$.
The maximum column density is 3.9$\times$ 10$^{19}$ cm$^{-2}$
at RA 23$^{\rm h}$ 07$^{\rm m}$ 52$^{\rm s}$, Dec 11\degree~31$'$ 28$''$.
(b) The first moment map of the MS VI region overlaid with
contours. Contours range from $-400$ to $-360$ \kms, with a step of 10
\kms. The grey-scale range is $-391$ to $-364$ \kms~with a linear transfer
function. Two prominent velocity ridges are shown with dashed lines: the
east ridge with a mean velocity of $-397$ \kms, indicated with
points labeled with A and B; the west velocity ridge with a mean
velocity of $-370$ \kms, indicated with points labeled with C, D and E. 
An increase in noise toward map edges is noticeable.}
\end{figure*}

The MS VI data is shown in Fig.~\ref{f:ms1_channels} where 
several velocity channels through the
spectral-line data cube are presented.
Fig.~\ref{f:ms1_posvel} shows position-velocity cuts. 
Fig.~\ref{f:ms1_0mom} shows maps of the  
HI column density and the intensity-weighted velocity
field (panels a and b). This region  has several
elongated clumps, inter-connected and forming  a  long curved
filament. Often, sub-clumps are found nested  inside larger clumps.
We identify individual clumps and describe
their properties in Section~\ref{s:clump_identification}.
Several interesting features appear in the data.

A large loop-like feature, centered
at  RA 23$^{\rm h}$ 07$^{\rm m}$ 26$^{\rm s}$, Dec 12\degree 47$'$ 30$''$, 
with radius of 50 arcmin,  
is visible between $v_{\rm LSR}=-360$ and $v_{\rm LSR}=-385$ \kms.  The
loop is present through $\sim$15 velocity channels but its size appears to
stay the same from channel to channel.  Although at first glance this
loop is reminiscent of HI holes associated with expanding shells of gas
seen in many galaxies,  because no systematic change  of the radius with
velocity is found this cannot be interpreted as an expanding shell. 
Velocity profiles along the loop are broad
but always single peaked. 

It is interesting to note that the most prominent clumps have a  dominant
orientation, from south-west to north-east, with  position  angle of $\sim
-40$\degree~(measured counter-clock wise from north).
This corresponds to a position angle of $\sim 90$\degree~in the
Magellanic coordinate system, defined by \cite{Wakker01} and is
parallel to the main axis of the MS \citep{Putman-thesis}.  Similar
orientation of clumps in MS VI was noted previously by  
Mirabel et al. (1979).

The total HI mass in the mapped region is 3.7$\times$ 10$^{5}$ M$_{\odot}$,
if the distance of 60 kpc is assumed, and it comprises 
about 28\% of the HI mass in the whole MS VI 
region estimated by \cite{Putman-thesis}.

The velocity profiles in this region are single-peaked but quite broad with
the dispersion being $\sim$ 15 \kms.
Several RA-velocity diagrams are shown in Fig.~\ref{f:ms1_posvel}.  At
about Dec 13\degree 05$'$ an interesting velocity gradient of almost 40
\kms~(LSR) is obvious.  This gradient
stays almost the same in the galactocentric velocity system, meaning
that a spatial variation of the projection of the LSR velocity relative
to the Galactic Center cannot explain
the velocity field in this relatively small, 40$\times$40 arcmin, field.

The velocity field shown in Fig. 4 is the first moment map of the HI
cube. The two prominent features in this map are north-south ridge-lines at
velocity $-365$ to $-375$ \kms~in the east, and velocity $-395$ to $-400$ 
\kms~in the west.  The points labeled A and B in 
Fig. 4 indicate the extent of
the east  ridge which may continue beyond the edge of our map. 
Points labeled C, D and E indicate the
extent of the  west ridge.
The steep velocity gradient evident in the middle panels of Fig. 3 is in
the  region bounded by points A, B, C and D in Fig. 4. This region alone
resembles the velocity pattern of a rotating disk, however the velocity
field of the  larger area does not fit this interpretation.  We discuss
this further in Section~\ref{s:alternative-interpretation}.

The velocity field in Fig. 4 is most likely caused by a combination of
blending of smaller clumps along the line of sight and  an interaction
between the MS clouds and the ambient medium.  With the Arecibo beam 
we are able to resolve several clumps blended together and forming a
velocity gradient, see Fig.~\ref{f:ms1_posvel}, especially around Dec
13\degree~08$'$.   Some of these clumps show tails in velocity and sharp
edges.  This may be due to interaction with the surrounding medium. We
discuss clump morphology in Section~\ref{s:clump-morphology}.  Mirabel et al.
(1979) saw a similar phenomenon in another MS VI region, while \cite{Wayte89}
attributed broad velocity profiles to the breaking of the MS clouds into
sub-clouds with different velocities.

\subsection{Morphology and velocity field of the MS V region}
 
\begin{figure*}
\epsscale{0.9}
\caption{\label{f:ms2_channels}{\bf [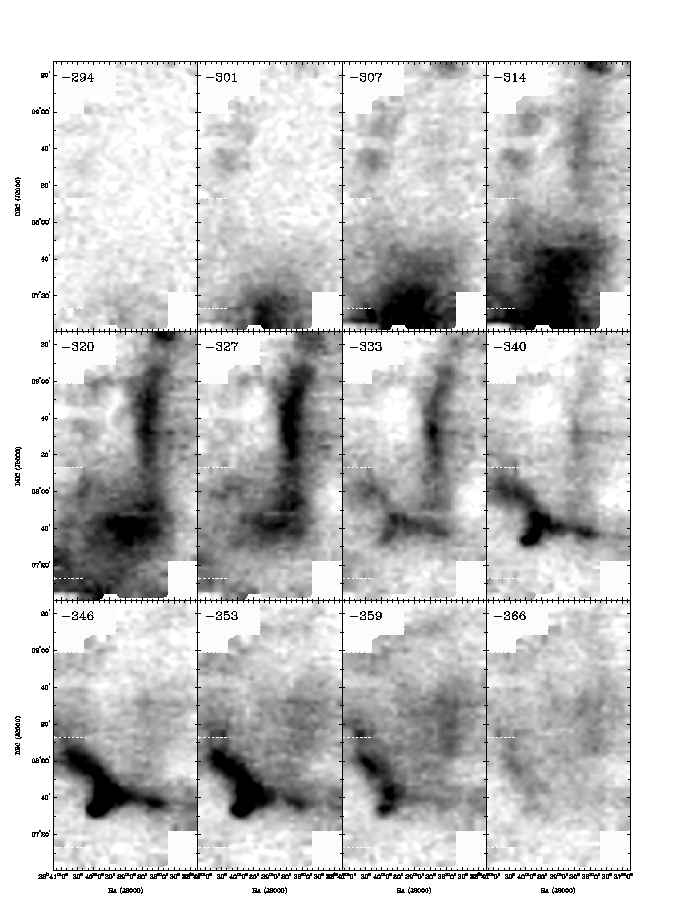]}The 
RA--Dec images of the MS V region for
different LSR velocities given in the top-left corner. The grey-scale
intensity range is $-0.01$ to 0.9 K with a linear transfer function. }
\end{figure*}

Velocity channel maps presented in Fig.~\ref{f:ms2_channels} show the general
morphology of the MS V region. Several dominant large-scale features  are
visible. In the $-320$ \kms~plane there is a round cloud 
centered around RA 23$^{\rm h}$ 39$^{\rm m}$, Dec
07\degree 35$'$ 00$''$.
This region shows single-peaked velocity profiles.
A long, straight filament stretches  north-south for almost one degree, at
RA 23$^{\rm h}$ 38$^{\rm m}$ 30$^{\rm s}$ from 
Dec 8\degree~to 9\degree,  at velocity of $\sim
-327$ \kms.  At higher velocities of $-340$ to $-355$ \kms, the most prominent
feature is a curved  filament,  centered at  RA 23$^{\rm h}$ 39$^{\rm m}$
30$^{\rm s}$,  Dec 07\degree 40$'$.   This consists of a core superposed on
an arc that extends for almost 1.3\degree.

\begin{figure*}
\epsscale{0.95}
\caption{\label{f:ms2_posvel}{\bf [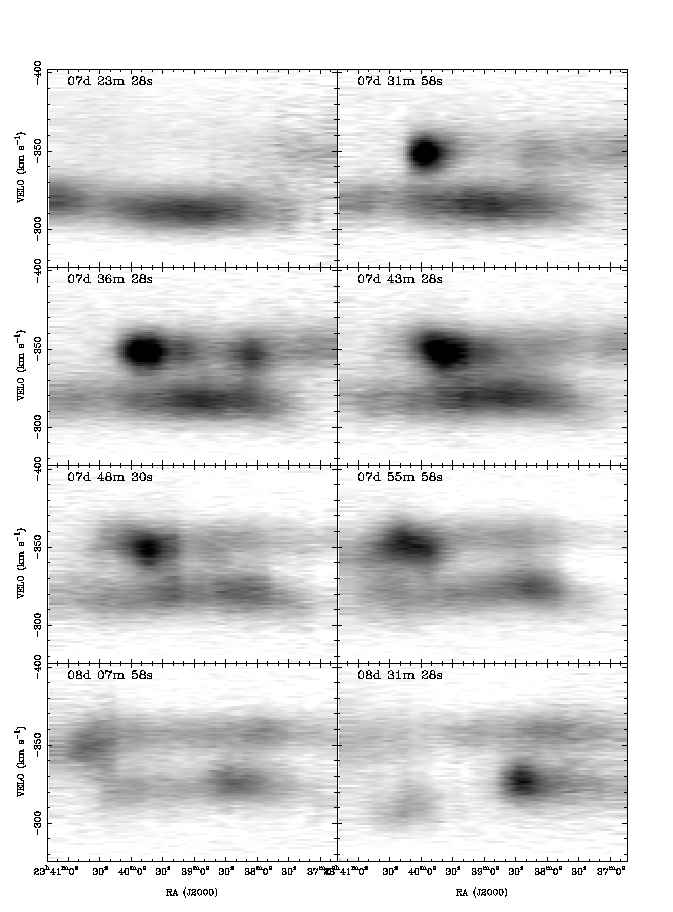]}Position--velocity 
diagrams of the MS V 
region taken at different Dec values,
given in the top-left corner. The grey-scale intensity range is 0 to 1.3 K 
with a linear transfer function.}
\end{figure*}

North of Dec = 07\degree 20$'$
all line profiles show double peaks.
This is shown in the position-velocity diagrams,
Fig.~\ref{f:ms2_posvel}, where a single velocity component is present
only in the upper left panel. All other panels show double profiles.
The two velocity peaks are roughly centered at $\sim -320$ and $\sim -360$
\kms~and run almost in parallel.

Double-peaked velocity structure has been seen previously in MS V by
Mirabel et al. (1979) at very similar central velocities.  \cite{Wayte89} 
also noted this velocity bifurcation in several regions in the MS. 
The most recent
HIPASS data, with very coarse velocity resolution of about 26 \kms, show
that spatially different filaments are common throughout the whole MS and
most likely have different origins \citep{Putman-thesis}.  Similar
line-splitting is found in several compact HVCs by \cite{Braun00}, who suggest 
that this structure is caused
by organized outflows or inflows.

What causes the dual velocity structure seen in MS V?  The velocity
separation between the two peaks of $\sim 40$ \kms~is reminiscent of
velocity features seen in the SMC.  This was formerly interpreted as a
signature of the SMC  being ripped into two clouds which began during its
last close encounter with the LMC some 200 Myr ago (Mathewson \& Ford
1984).  Recently, high spatial and velocity resolution observations of the
SMC \citep{Staveleyetal97,Stanimirovic99} discovered more
complex velocity profiles and numerous expanding shells of gas. This
line-splitting in the SMC is now attributed primarily to the effect 
of expanding
shells. Multiple structures in velocity are also  seen in the Magellanic
Bridge region \citep{McGeeNewton86}.  No evidence for shells has been
found in the MS.

We have fitted overlapping Gaussians to the blended velocity components,
and the
results show that their velocity separation is quite constant over the area
of MS V, as  seen in Fig. 6.

\subsubsection{Gaussian decomposition}

\begin{figure*}
\epsscale{1.0}
\plotone{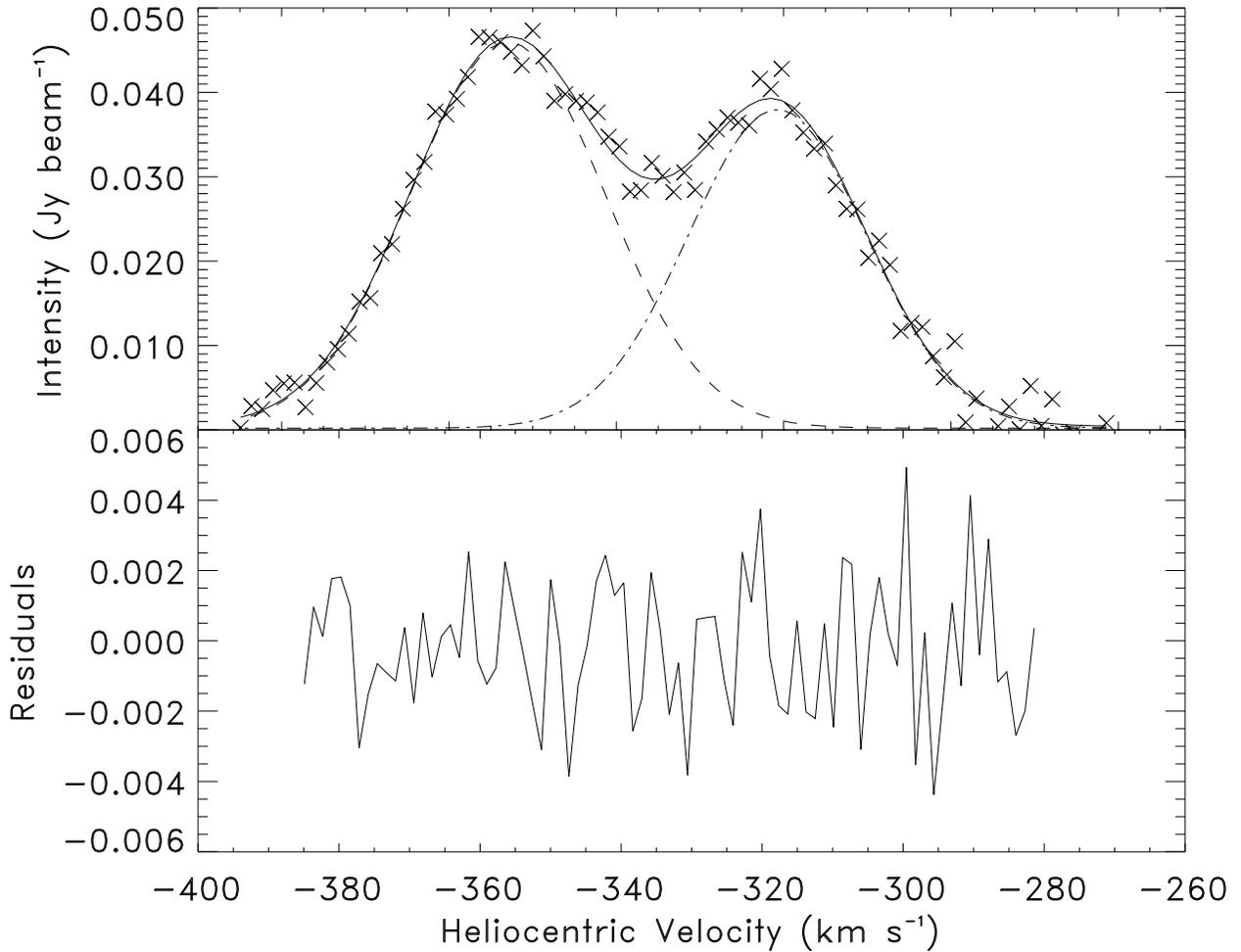}
\vspace{1cm}
\caption{\label{f:example_fit}An example of Gaussian decomposition at RA
23$^{\rm h}$ 39$^{\rm m}$ 46$^{\rm s}$, Dec 08\degree 01$'$ 00$''$. Two
different fitted components are shown as dashed and dot-dashed lines,
while their sum is shown as a solid line.}
\end{figure*}

\begin{figure*}
\epsscale{1.0} 
\plotone{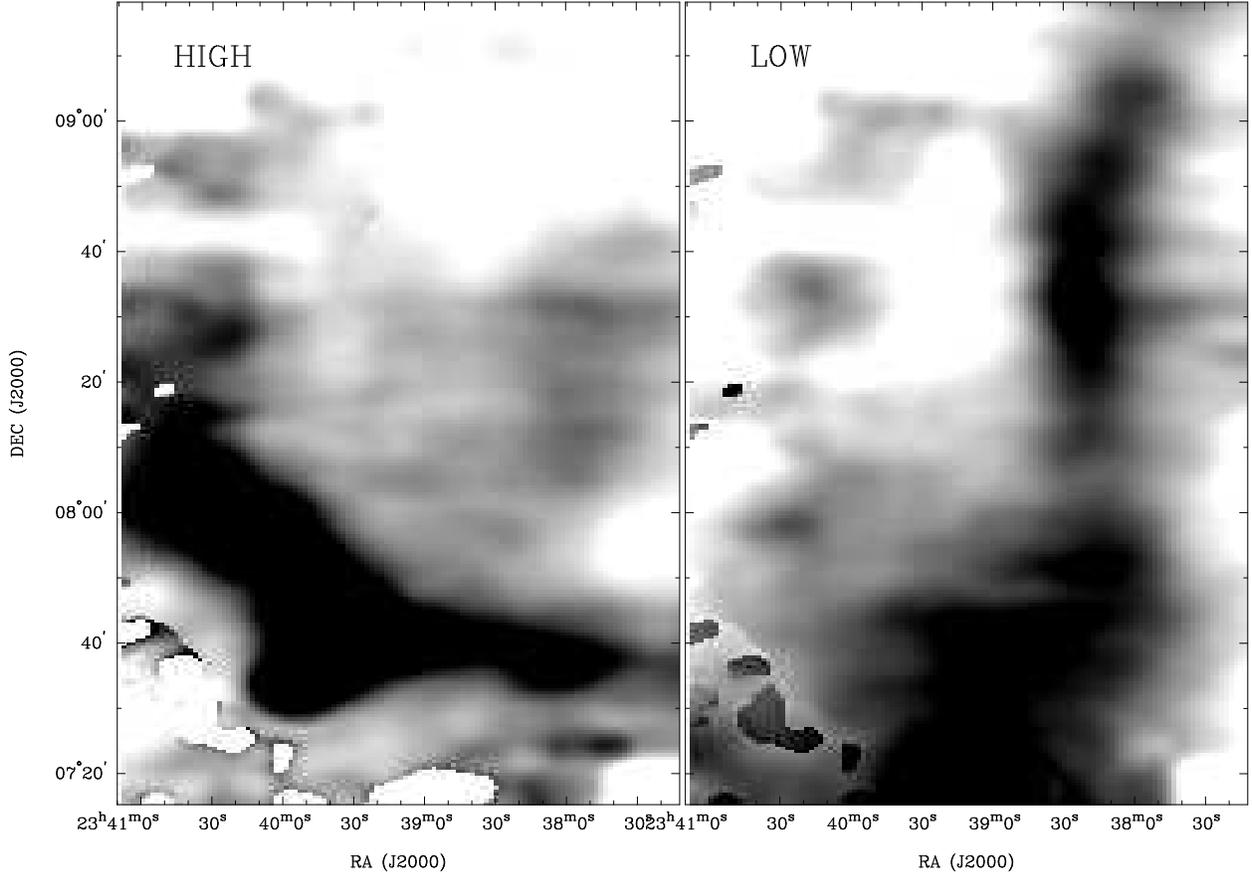}
\caption{\label{f:6planes_1}An integrated 
intensity image of two Gaussian
components fitted to MS V velocity profiles. The grey-scale intensity range
is 1 to 4 $\times10^{19}$ cm$^{-2}$ for the higher velocity component, and
1.2 to 4 $\times10^{19}$ cm$^{-2}$ for the lower velocity component.}
\end{figure*}

To model the velocity profiles with one or a superposition of two Gaussian
functions, the data cube was binned spatially in squares of 5 by 5 pixels.
As an example, Fig.~\ref{f:example_fit} shows one velocity profile, its
decomposition into two Gaussian functions, and the residuals after
subtracting the fit from the data.   The mean rms value of the residuals in
the whole region is close to  the noise level, with its maximum below the
2-$\sigma$ level.  The two separate components are centered around $-350$
\kms, the `higher' velocity component (`H'), and $-320$ \kms, the `lower'
velocity component (`L').
The integrated intensity for each velocity component is shown in
Fig.~\ref{f:6planes_1}. The higher velocity component is the core-arc
feature described above, plus a diffuse feature to the north-west.  The lower
velocity component is the round cloud and the north-south filament.  The HI
mass of the higher velocity component is 2.3$\times 10^{5}$
M$_{\odot}$. The HI mass of the lower velocity component is 3.2$\times
10^{5}$  M$_{\odot}$ (if the distance of 60 kpc is assumed).  
These together make up only 15\% of the  HI mass of
the entire MS V region, 3.9$\times 10^{6}$ M$_{\odot}$ \citep{Putman-thesis}.

A slight gradient in velocity of about 10 \kms~is seen from the north to
the south and is shown by the slight tilt  of the ridge lines in Fig. 6. These
trace roughly   $\sim -360$ to $\sim -350$ \kms~for the higher velocity
component, and $\sim -325$ to $\sim -315$ \kms~for the lower velocity
component, from right to left on the lower middle  panels of Fig. 6.   The
dispersion of the lower velocity component has everywhere a value of about
10 \kms. The dispersion of the higher velocity component shows a
significant variation with position. In the north,  the velocity profiles
are very narrow, with typical velocity dispersion of about 6 \kms. The
dispersion increases towards the south reaching about 13 \kms~along the
core-arc feature.

We have experimented with fitting more than two velocity components to the
data, however the resultant fits did not improve the residuals
significantly.   Neither MS V nor MS VI regions show narrow line components
indicative of cold HI (with temperature of 30 -- 300 K). This is in
agreement with theoretical predictions by \cite{Wolfire95}, that no cold
cores should be found in clouds in the Galactic Halo at heights  above 20
kpc.

\section{Clump analyses}
\label{s:clump-identification+analyses}

\subsection{Clump identification}
\label{s:clump_identification}

\begin{figure*}
\epsscale{0.9}
\caption{\label{f:msV_clumps_new}{\bf [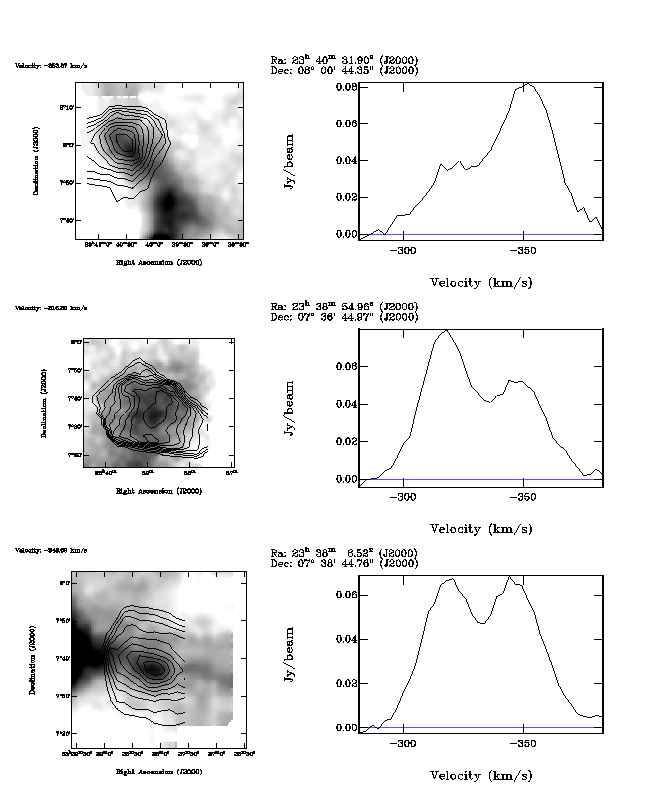]}Three 
fully mapped and well isolated
clumps in the MS V region. For each clump, 
the left-hand side shows an HI intensity image
overlaid with contours of the total column density, while the right-hand
side shows a velocity profile at the clump's center.}
\end{figure*}

\begin{figure*}
\epsscale{0.75}
\caption{\label{f:msVI_clumps_new}{\bf 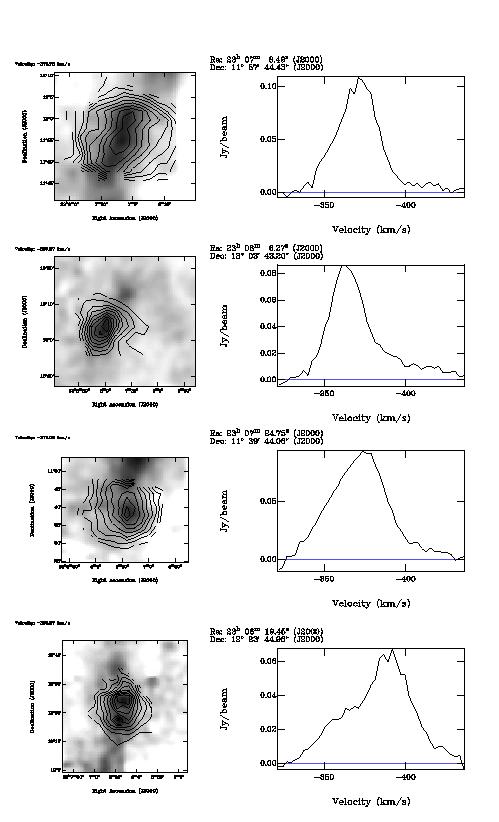]}
Four fully mapped and well isolated
clumps in the MS VI region, from the top to the bottom. For each clump, 
the left-hand side shows an HI intensity image
overlaid with contours of the total column density, while the right-hand
side shows a velocity profile at the clump's center.}
\end{figure*}

\begin{table*}
\caption{\label{t:clumps_table}A summary of clump properties.}
\begin{tabular}{lcccccc}
\noalign{\smallskip} \hline \hline \noalign{\smallskip}
Clump ID & RA (J2000) & Dec (J2000) & LSR Velocity & Radius & FWHM &
$T_{\rm peak}$\\
 & (hh:mm:ss) & (dd:mm:ss) & (\kms) & (arcmin) & (\kms) & (K)\\
\noalign{\smallskip} \hline \noalign{\smallskip}
Small clumps & & & & & & \\
\noalign{\smallskip} \hline \noalign{\smallskip}
MS V: & & & & & & \\
   1&   23:40:23&     07:58:45& $-354.5$&   13&      21.1&
0.8\\
   2&   23:38:55&     07:38:45& $-318.3$&   18&      18.6&
0.7\\
   3&   23:38:07&     07:36:45& $-344.2$&   15&      24.2&
0.7\\

MS VI: & & & & & & \\
 
  1&   23:07:07&     11:55:45& $-371.1$&   13&      20.7&
1.0\\
  2&   23:07:55&     13:03:45& $-363.3$&   11&      16.7&
0.8\\
  3&   23:07:23&     11:37:45& $-373.7$&   12&      16.3&
0.8\\
  4&   23:06:19&     12:23:45& $-391.8$&   13  &    16.4 &
0.6\\
\noalign{\smallskip} \hline \noalign{\smallskip}
Large clumps & & & & & & \\
\noalign{\smallskip} \hline \noalign{\smallskip}
MS VH & 23:40 & 08:10 &$\sim -320$  & 64$\times$28& 23.5 & 0.9 \\
MS VL & 23:39& 08:19&$\sim -350$ & 54$\times$30 &20.6 & 1.2\\ 
MS VI & 23:07 & 12:22&$\sim -370$ & 65$\times$31 & 35.5& 1.7\\

\noalign{\smallskip} \hline \noalign{\smallskip}
\end{tabular}
\end{table*}

In order to study the properties of the MS clumps and the indirect clues
they can provide about the properties of the Galactic Halo,  we have
attempted to identify individual clumps. We used two different approaches.\\

(i) To identify smaller clumps and sub-clumps within larger clumps, the
three-dimensional clump-finding algorithm, `clfind' coded in IDL 
\citep{Williams94} was applied to both data sets.
A spatial binning by 4 pixels, and a velocity binning by 2 pixels was
applied to reduce the  computing time. `Clfind' views data as a set of
contour levels. It locates local maxima and defines clumps by tracing them
by connecting pixels, at each contour level, that are within one resolution
element of each other. The algorithm does not assume any particular
geometry for the clumps.  The most important input parameter for `clfind'
is the contour increment. This was set to twice the noise level, as
recommended in Williams et al. (1994). All selected clumps were  further
vigorously inspected by eye  by overlaying the `clfind' output on the
original data using the {\sc karma} visualization package \citep{KarmaRef}.
Only a few (seven) clumps that are isolated and fully inside the edges of
our map  were considered for further analysis.
Those clumps, three in MS V and four in MS VI, are shown in
Figs.~\ref{f:msV_clumps_new} and \ref{f:msVI_clumps_new}.  For each
clump the central position, central velocity,  clump size ($R_{\rm c}$)
measured from the integrated intensity map, FWHM velocity line width
($\Delta v$), and peak brightness temperature were estimated and are listed
in Table~\ref{t:clumps_table}.

(ii) The clumps found by `clfind' may or may not be distinct 
physical structures. On larger scales, all of MS VI and the `L' 
and `H' components of MS V could be considered as clumps themselves.
The corresponding quantities for these three largest clumps are also listed in
Table~\ref{t:clumps_table}.

\subsection{Observed and derived properties}
\label{s:cloud-properties}

\begin{figure*}
\epsscale{1.0}
\plotone{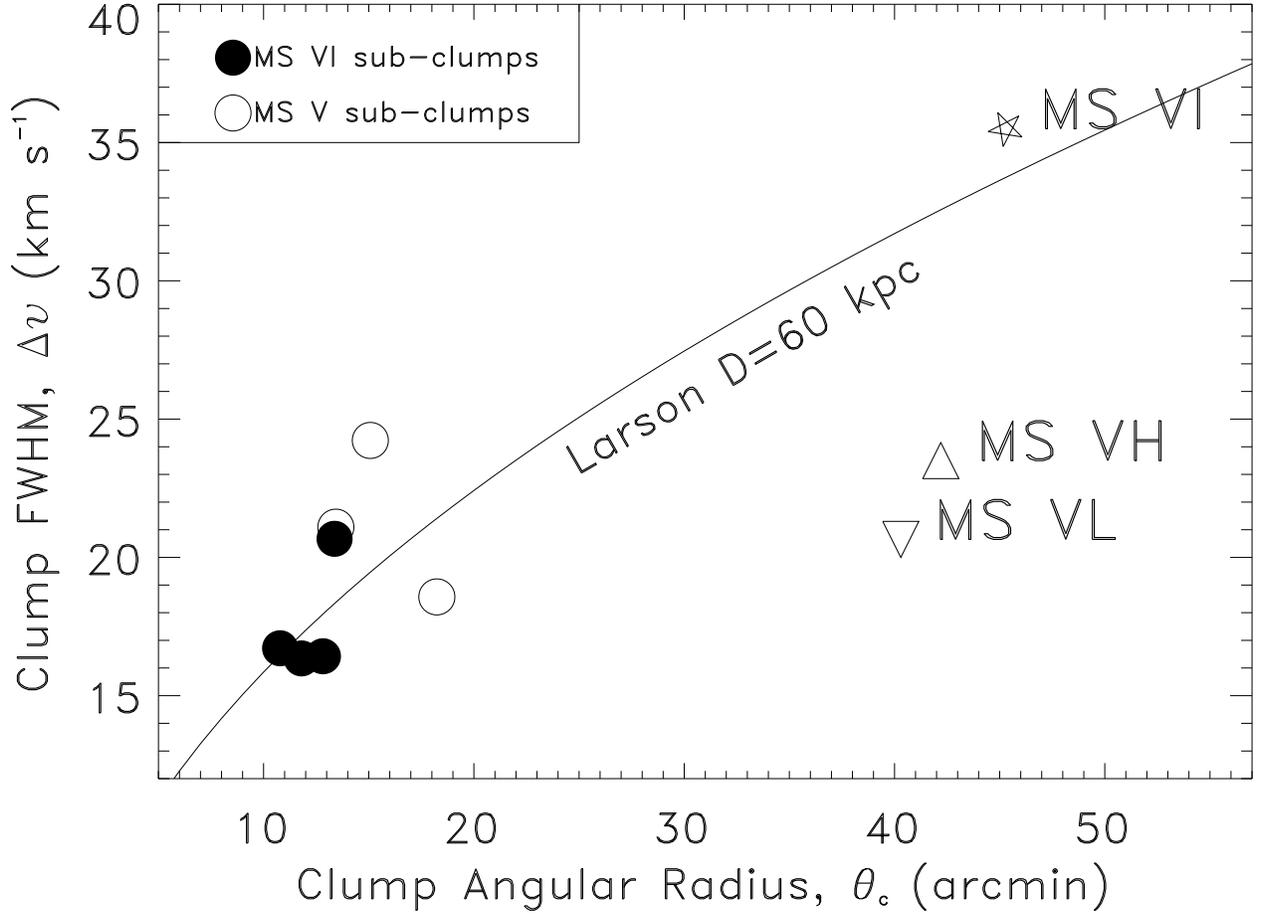}
\caption{\label{f:clumps_all_rad_fwhm}Clump angular radius 
($\theta_{\rm c}$) vs FWHM ($\Delta v$) for clumps in
the MS VI region (filled circles), and in the MS V region (open
circles). The whole MS VI region is shown as a star, and MS V `L' and MS V
`H' components with downward and upward pointing triangles, respectively. 
Larson's scaling law for the line width and cloud size, 
$\Delta v \propto \left [ D \times \tan \left (\theta_{\rm c} \right) 
\right ]^{0.5}$, is overlaid for the assumed distance $D=60$ kpc.}
\end{figure*}

Fig.~\ref{f:clumps_all_rad_fwhm} shows the angular size and line
width for all the clumps. 
A correlation between line width and cloud size has 
been found for interstellar clouds in a
number of  surveys (references in Vazquez-Semadeni et al. 1997). 
This is one of the famous Larson scaling laws: $\Delta v \propto
R_{\rm c}^{\beta}$. For the case of
self-gravitating molecular clouds $\beta=0.5$ \citep{Myers88}, 
a  slightly flatter
slope of 0.4 was found by \cite{Falgarone92} in molecular
clouds that are very little contaminated by star formation.  
We overlay the Larson relation in
Fig.~\ref{f:clumps_all_rad_fwhm} for the assumed distance of 60 kpc. 
While all sub-clumps and one of the largest clumps (the whole of MS VI) 
appear to  follow the correlation, two other large clumps have $\Delta v$
lower by about 10 \kms~than the predicted values.

We now derive several important clump properties.
These properties depend on the assumed distance of the MS. 
Figs. 12 through 16 have double axes  corresponding to the extreme distances of
60 and 20 kpc (see Section~\ref{s:theories}). 
For all these figures, the left-hand and bottom axes 
correspond to the far distance, and the right-hand and top axes correspond to 
the near distance.

\begin{figure*}
\epsscale{1.0}
\plotone{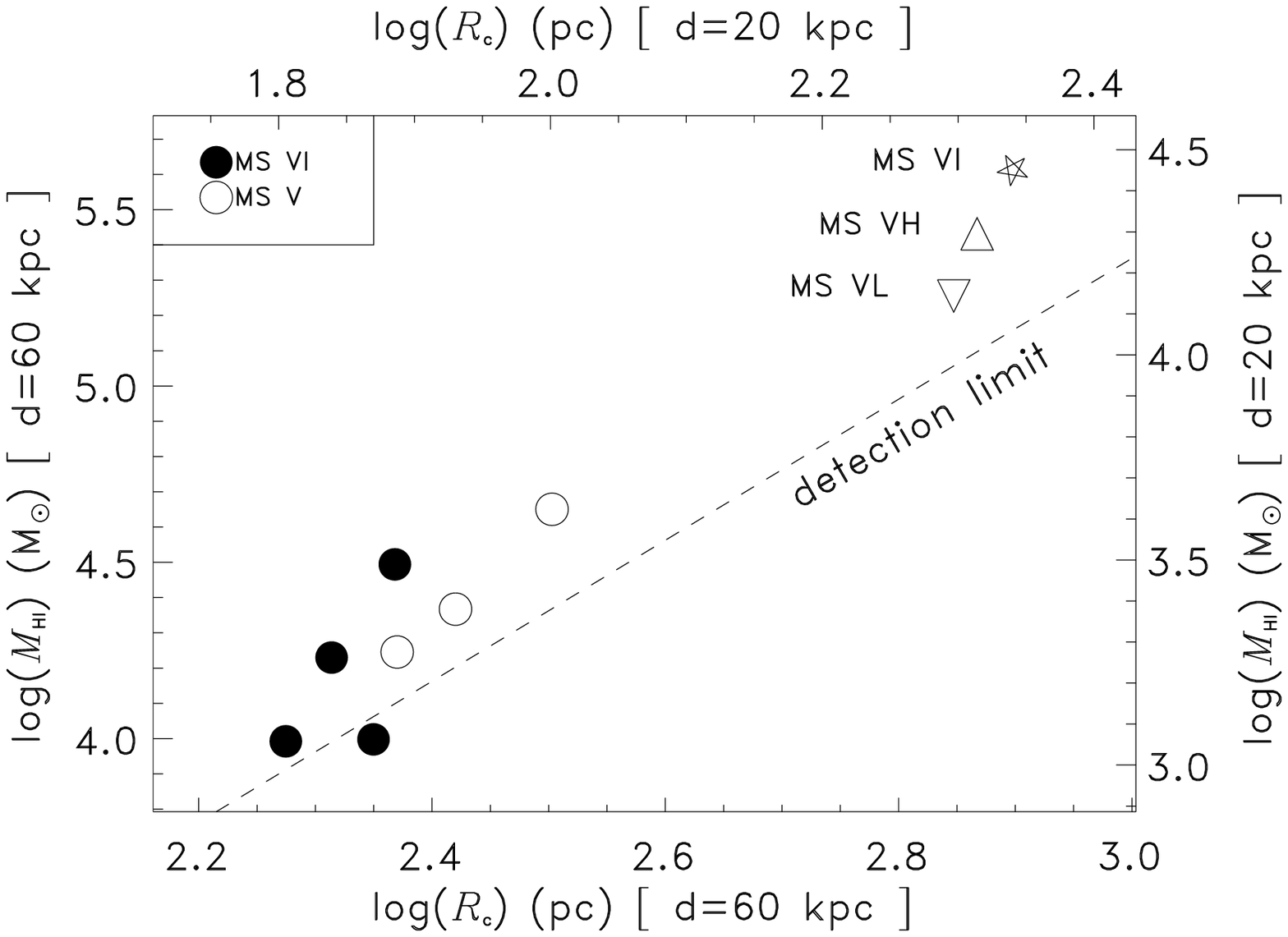}
\caption{\label{f:clump_rad_mass} The total HI mass ($M_{\rm HI}$) as a
function of cloud size ($R_{\rm c}$) on a log--log scale. 
As previously, clumps in
the MS VI region are shown as filled circles, and in the MS V region as open
circles. The whole MS VI region is shown as a star, and MS V `L' and MS V
`H' components with downward and upward pointing triangles, respectively.  The
dashed line shows our 3-$\sigma$ detection limit. The left-hand and bottom
axes correspond to values derived assuming distance of 60 kpc, the
right-hand and top axes correspond to the near distance of 20 kpc.}
\end{figure*}

The total HI mass is derived by integrating the brightness temperature over the
entire line width. The HI mass is plotted as a function of cloud size
in Fig.~\ref{f:clump_rad_mass}.  
For a given size our noise level sets a minimum detectable cloud mass.
Using the narrowest measured clump line widths of 
$\sim$ 10 \kms~ we get  the threshold indicated by the dashed line on
Fig.~\ref{f:clump_rad_mass}. 
Fig.~\ref{f:clump_rad_mass} suggests that there may be many 
more structures like our clumps
in MS V and MS VI which could be found by a more sensitive survey.

\begin{figure*}
\epsscale{1.0}
\plotone{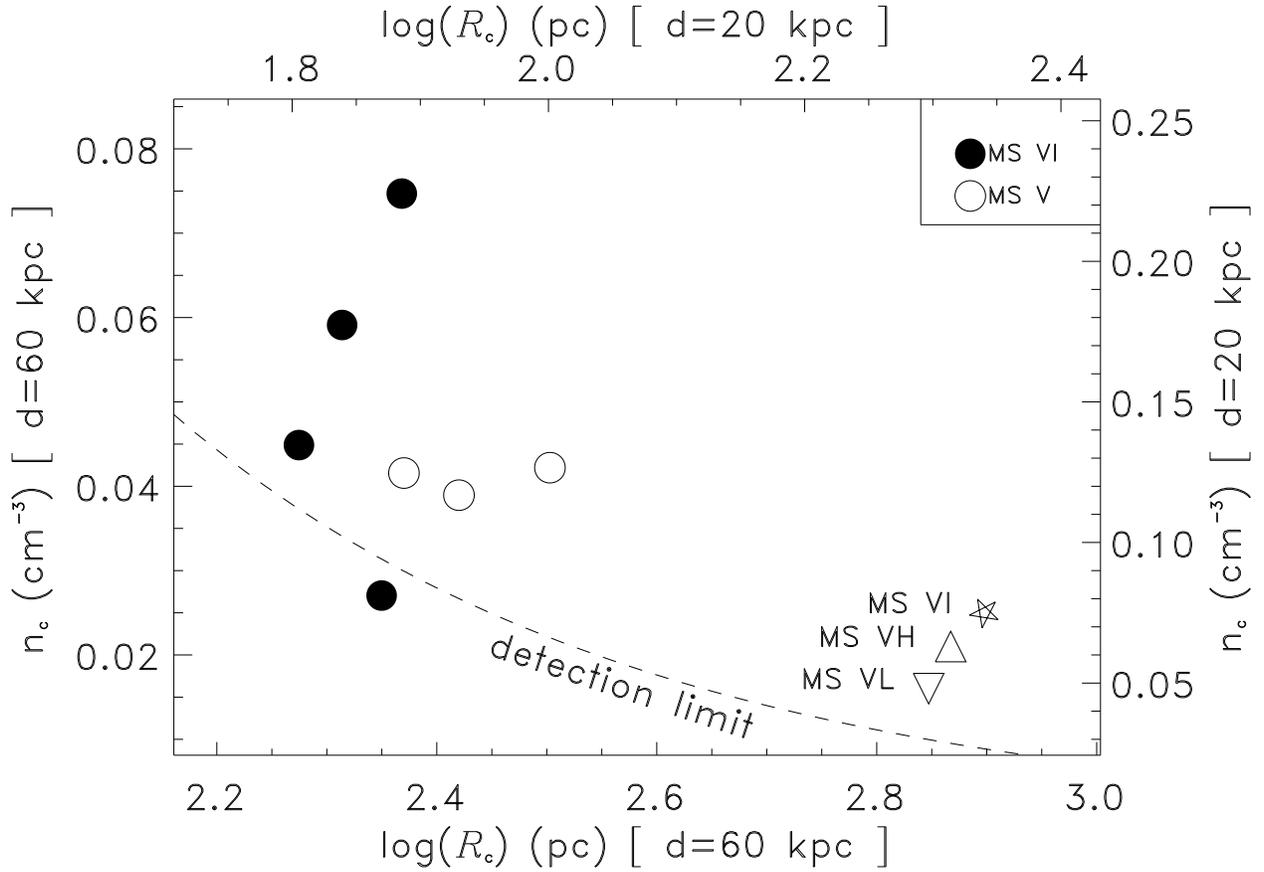}
\caption{\label{f:clump_size_rho}The average HI volume density ($n_{\rm c}$)
as a function of cloud size ($R_{\rm c}$).
The symbols and axes have the same meanings as in the preceding plots. 
The dashed line represents the 3-$\sigma$ detection limit.}
\end{figure*}

Assuming that clouds are spherically symmetric, using their HI mass and radius
we can estimate an average HI volume density, $n_{\rm c}$. 
We find the values of $n_{\rm c}$ shown in Fig.~\ref{f:clump_size_rho}, 
where the 3-$\sigma$ detection threshold is again shown by the dashed line.
A typical value for the small clumps is  
0.05 cm$^{-3}$ and for the large clumps 0.02 cm$^{-3}$, for the 60 kpc 
distance assumption. For the 20 kpc distance
the volume density increases by a factor of three, as shown by the right-hand 
scale on Fig.~\ref{f:clump_size_rho}.

\section{Discussion}
\label{s:discussion}

\subsection{Cloud confinement}
\label{s:cloud-stability}

We now turn to cloud stability issues and the survival of the MS 
so far away from the MCs. There are several questions to ask about 
the HI structures which we find.\\

\begin{figure*}
\epsscale{1.0}
\plotone{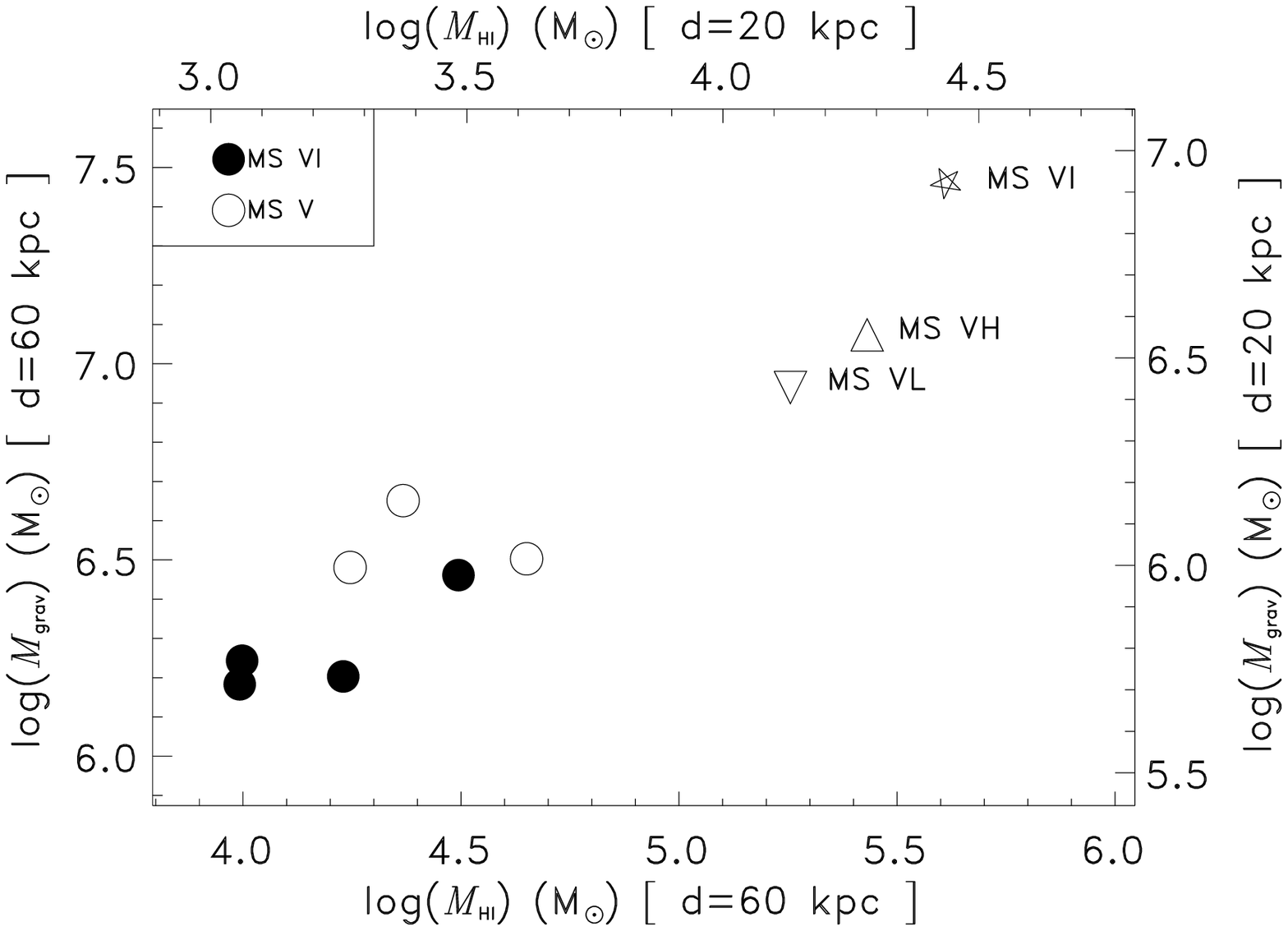}
\caption{\label{f:clump_massHI_massGrav}The hypothetical gravitational mass
($M_{\rm grav}$) as a function of measured HI mass ($M_{\rm HI}$) on a
log--log scale.} 
\end{figure*}

(1) {\it Are the MS clumps gravitationally confined?}\\ 
In order
to be gravitationally confined, a clump's escape  velocity must be larger
than $\Delta v$/2. This requires that the clump's  total mass is $M_{\rm
grav} \geq R \Delta v^{2}/8G $.
Fig.~\ref{f:clump_massHI_massGrav} compares the measured HI masses of our
clumps with this hypothetical gravitational mass.  
For all our points $30<M_{\rm grav}/M_{\rm HI}<250$, if the distance of 60 kpc
is assumed, and $120<M_{\rm grav}/M_{\rm HI}<750$, for the 20 kpc distance
assumption.
Thus
gravitational confinement would require unreasonable amounts of dark matter.

\begin{figure*}
\epsscale{1.0}
\plotone{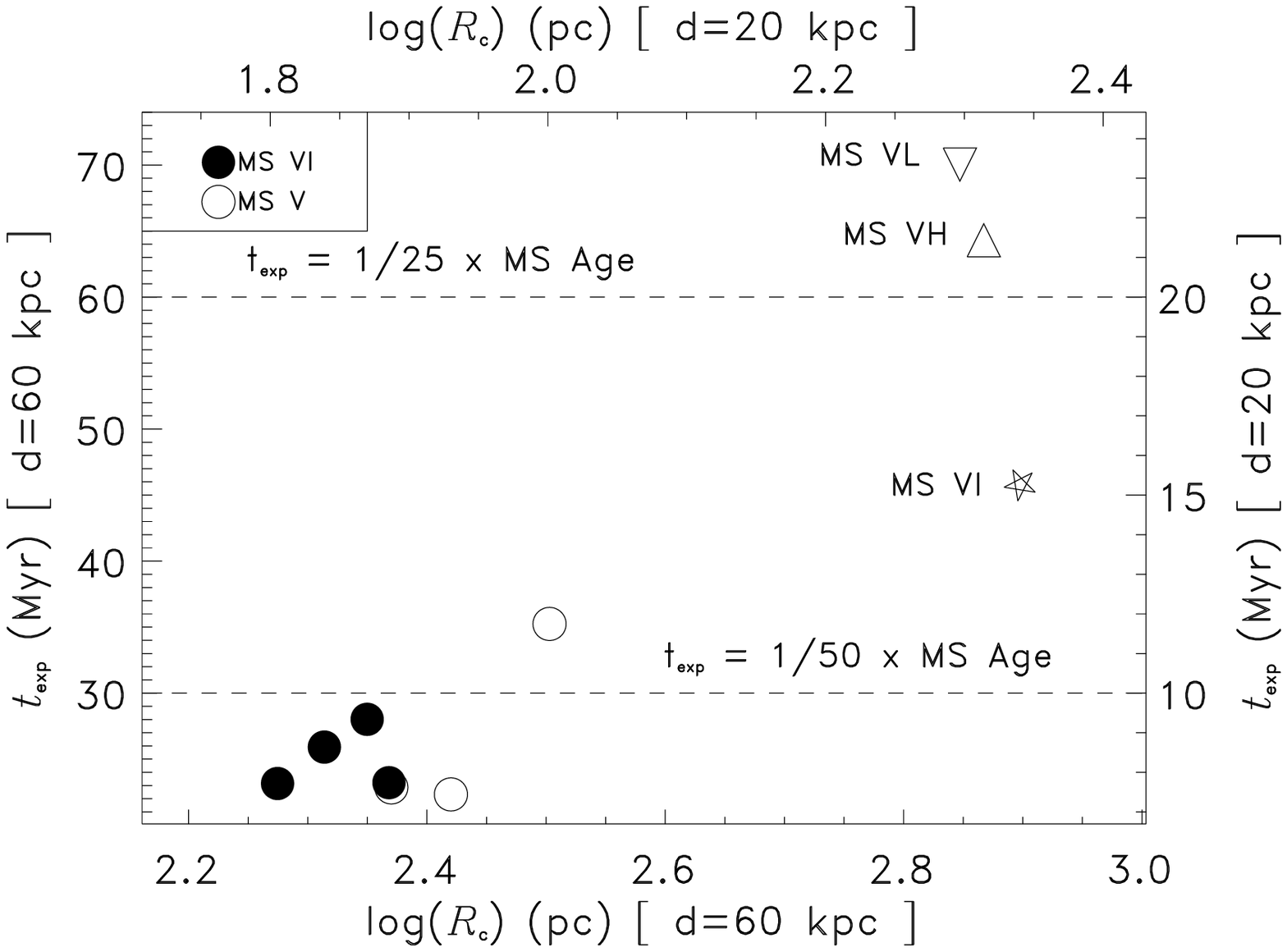}
\caption{\label{f:clump_rad_age}Cloud expansion age ($t_{\rm exp}$) 
as a function of cloud size ($R_{\rm c}$). 
Dashed lines show $t_{\rm exp}$ as a fraction of the MS age. }
\end{figure*}

(2) {\it Are the MS clumps unbound and freely expanding?}\\ 
In the tidal
model the MS age is 1.5 Gyr, while in the ram pressure model the age is 500
Myr.  If the clouds are freely expanding their expansion time is given by:
$t_{\rm exp}=2R/{\Delta v}$. These numbers are much less than the MS age.
Fig.~\ref{f:clump_rad_age} plots the expansion age as a function of clump
radius.  The expansion age as a fraction of the MS age   is indicated by
the horizontal lines which apply to either model.  This huge discrepancy
between the expansion time and the age  is the fundamental  problem of the
gas dynamics of the MS, as pointed out first by Mirabel et al. (1979).
Barring the unreasonable dark matter halos required for gravitational
confinement the simplest explanation is that  the MS clouds are confined by
a hot gaseous  Galactic Halo.

\subsection{Density of the Milky Way Halo}
\label{s:halo-density}

If the MS is confined by the Galactic Halo, then the external pressure, 
$P_{\rm h}=kn_{\rm h}T_{\rm h}$, 
must equal the clouds' 
internal pressure $P_{\rm c}=kn_{\rm c}T_{\rm c}$: 
$P_{\rm c} \leq P_{\rm h}$. The relevant value of $T_{\rm c}$ is not 
the microscopic 
kinetic temperature but the total random motion of the atoms which we
describe as a temperature defined by  the line-width: 

\begin{equation}
T_{\rm c}= \frac{\Delta v^{2}}{8 \ln 2} \frac{k}{m_{\rm H}}=
21.9~{\rm K} \times \left( \frac{\Delta v}{\rm km~s^{-1}} \right)^2.
\end{equation}

\begin{figure*}
\epsscale{1.0}
\plotone{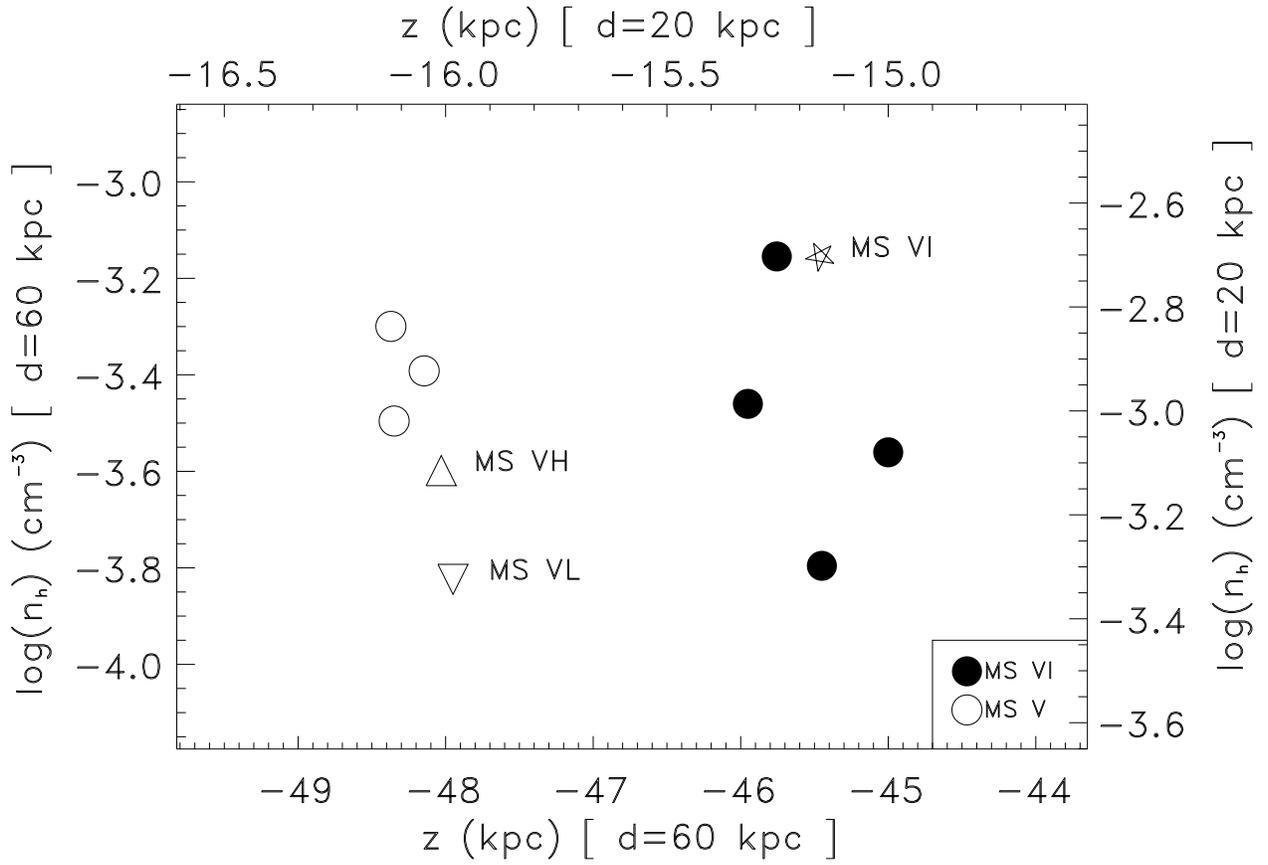}
\caption{\label{f:rho_lower}The logarithm of the Galactic Halo 
density ($n_{\rm h}$) as a function of distance from the Galactic 
Plane ($z$). }
\end{figure*}

Assuming a halo temperature of 10$^{6}$ K and using $\Delta v$ for the clumps
in Table 2 and the densities $n_{\rm c}$ illustrated on 
Fig.~\ref{f:clump_size_rho}, we get Halo densities $ n_{\rm h}$
shown in Fig.~\ref{f:rho_lower} as a function of distance
from the Galactic Plane, $z$.
Typical values for the near distance ($z \approx 15$ kpc) 
are $n_{\rm h} = 10^{-3}$ cm$^{-3}$ and for the far distance ($z \approx 
45$ kpc) are $n_{\rm h} = 3 \times 10^{-4}$ cm$^{-3}$.
Note that these estimates are upper limits on the Halo density as 
magnetic, turbulent or ram pressure contributions were not included.

\subsection{Expected Halo density}

Current theories of galaxy formation predict the existence of a hot 
Galactic Halo containing the leftover gas which was unable to 
cool since the formation 
of the Galaxy. However, the density structure of the Halo is 
still very uncertain, especially at large distances from the Galactic Plane. 
Here we compare our results with several previous estimates, 
both observational and theoretical.

\subsubsection{Observational constraints}

The only direct observations of the hot Halo gas come from 
measurements of the soft X-ray
background \citep{Snowden97}. These X-ray maps reveal
hot gas with a temperature $\sim 10^{6.6}$ K, a scale-height of 1.9 kpc and
an electron density of $\sim$ 3.5$\times 10^{-3}$ cm$^{-3}$. 
Indirect observational constraints
on the Halo density are more common. These are based on 
dispersion measure measurements of pulsars
in globular clusters, or on the observational properties of HVCs, 
including the MS clouds.

Mirabel et al. (1979) estimate that $n_{\rm h}=7\times10^{-4}$ --
$7\times10^{-5}$ cm$^{-3}$ for the distance range 10 -- 100 kpc from HI
observations of the MS and under similar confinement considerations 
as discussed in this paper.
Similar values, $n_{\rm h} \geq 10^{-4}$ cm$^{-3}$, were suggested from
H$_{\alpha}$ observations along the MS by \cite{Weiner96}. 
\cite{Quilis01} applied the terminal velocity model by \cite{Benjamin97} 
for HVCs
moving at $\sim300$ \kms, at distances 10 -- 100 kpc and with 
HI column density of $10^{20}$
cm$^{-2}$. This results in the requirement that $n_{\rm
h}(10~{\rm kpc})<3\times10^{-3}$ cm$^{-3}$ and $n_{\rm
h}(100~{\rm kpc})<3\times10^{-4}$ cm$^{-3}$, under the assumption 
that these are pure gas clouds.
Our results agree well with the estimates from these three studies.

Another constraint for the Halo density at the distance of 130 kpc 
of $\sim 10^{-5}$ cm$^{-3}$ comes from
HI observations of HVC 125$+$41$-$207 by \cite{Bruns01}.
A significantly lower value for the upper limit on the Halo density, 
$n_{\rm h}(50~{\rm kpc})<10^{-5}$ cm$^{-3}$, was imposed by
\cite{Murali00} by allowing the whole MS VI cloud to survive at the tip of the
MS for about 500 Myr. We look more closely into this approach in 
Section~\ref{s:approaches}.

\subsubsection{Theoretical predictions}

\cite{Benjamin97} considers a Galactic Halo  
consisting of three components: the warm
ionized layer of HII parameterized by \cite{Reynolds93}; the 
mean HI density distribution given by \cite{Dickey90}; 
and the hot Halo component with $T_{\rm
h}=10^{6}$ K prescribed by Wolfire et al. (1995).
While the first two components were parameterized observationally, the third
component was constructed theoretically 
assuming an idealized isothermal halo in hydrostatic
equilibrium, with the mid-plane
density matching X-ray emission data by \cite{Garmire92}. 
The combination of all three components results in 
the analytic representation for $n_{\rm h}$ as a function of $z$ 
given by Benjamin \& Danly (1997):
\begin{equation}
n_{\rm h}(z)=1.1\times10^{-3} \left [ 1+\left (z^{2}_{\rm kpc}/19.6 \right) 
\right ]^{-1.35} {\rm cm}^{-3}.
\end{equation}
At $z=15$ kpc the expected density is $n_{\rm h}=4\times10^{-5}$ 
cm$^{-3}$, and at
$z=45$ kpc $n_{\rm h}=2\times10^{-6}$ cm$^{-3}$. 
These values are 25 -- 150 times lower than those
found in Section~\ref{s:halo-density}.

The ram-pressure model for the MS origin by Moore \& Davis (1994) requires
slightly higher values for the Halo density in order
to match the kinematics of MS clouds: 
\begin{equation}
n_{\rm h} = 0.002/ \left[1 + \left(R/12~{\rm kpc} \right) \right]^2~{\rm 
cm^{-3}}.
\end{equation}
Hence, at a distance of 20 kpc this yields a gaseous halo with 
$n_{\rm h}=3\times10^{-4}$ cm$^{-3}$, while at a distance of 60 kpc 
$n_{\rm h}=6\times 10^{-5}$ cm$^{-3}$.
Our estimates are much closer to the values predicted by this model 
but still several times higher.

\subsection{Nature of Galactic infall}
\label{s:approaches}

There are (at least) three different approaches present in the 
literature concerning the dynamics of clouds moving through 
an ambient medium. As understanding of the nature of
this interaction can in turn
provide clues about the properties of the Halo, here we 
discuss these approaches briefly.\\

(1) {\it The mass loading approach} \\
As the MS moves through the hot ambient Halo gas its clouds are 
subject to a drag force. Murali (2000) showed that this motion 
is actually dominated not by drag but by strong heating from accretion. 
If the Halo density is too high, heating due to accretion can cause
cloud evaporation.
This process is usually referred to as mass-loaded flow and it is common in
comet interactions with the solar wind \citep{Flammer91}. 
The Halo gas ionizes the cloud's leading edges and picks up the `new material'
in a very asymmetric way that is 
constrained by the magnetic field: magnetic field lines slip 
past the cloud therefore allowing ablation of material only from the poles. 
Murali (2000) accounted for these effects and estimated that 
for the survival of the whole MS IV region for 500
Myr, at a distance of 50 kpc, $n_{\rm h} < 10^{-5}$ cm$^{-3}$ is
required. The key factor in this approach
is the cooling resulting from a cloud's mass-loss.

Following Murali's calculations (their equation 1) for the MS V and MS 
VI clouds studied here, drag and accretion set a very low upper limit for 
the Halo gas density: $n_{\rm h}\sim 10^{-7}$ cm$^{-3}$ at $z=45$ kpc, and 
$n_{\rm h}\sim 10^{-6}$ cm$^{-3}$ at $z=15$ kpc, if mass-loss 
is not included. 
These values are significantly lower than theoretical predictions.
Unless a significant cooling due to mass-loss is introduced, it is very
hard for the hot Halo gas to exist in this model.

If the mass-loss rate, \.{M}$=2.5\times10^{-2}$ M$_{\odot}$ 
yr$^{-1}$ (Murali 2000)  is included, the requirements are 
comparable to our estimates: 
$n_{\rm h}\sim 3\times 10^{-4}$ cm$^{-3}$ at $z=45$ kpc and
$n_{\rm h}\sim 3\times 10^{-3}$ cm$^{-3}$ at $z=15$ kpc.
A slightly lower rate of \.{M}$=10^{-3}$ M$_{\odot}$ yr$^{-1}$ 
results in ten times lower Halo densities. 
However, the mass-loss rates of \.{M}$=10^{-3}$ or few$\times10^{-2}$ 
M$_{\odot}$ yr$^{-1}$ throughout the whole cloud life-time are 
too high in comparison with the cloud HI masses discussed in
Section~\ref{s:cloud-properties}. Such mass-loss rates would imply that cloud
HI masses were 10 -- 1000 times higher at the time of the MS formation.
Smaller rates that are in 
agreement with the current HI masses are not sufficient to allow the 
existence of the Halo gas. Detailed simulations of the mass-loaded
mechanism may help to constrain parameters such as mass-loss rate. \\

(2) {\it Ballistic approach}\\
\cite{Gregori99} showed results of a 3-D study of a
ballistic interaction of a moderately supersonic dense cloud with a
warm magnetized medium. Since
clouds are supersonic, their motion leads to the formation of a forward bow
shock and a reverse crushing shock propagating through the cloud. As the
cloud moves, its surface is subject to several instability mechanisms, the
most disruptive of these being the Rayleigh-Taylor instability (R-T) 
that develops at the interface between two fluids when the lighter 
fluid accelerates the heavier one.  These
instabilities slowly disrupt the entire cloud and can dramatically change
its morphology. 
In contrast to the mass-loaded flows, magnetic field lines here 
stay trapped, causing the development of strong
magnetic pressure at the leading edge of the cloud.

The time-scale over which this dramatic change occurs is several times
the so called `crushing time', $\tau_{\rm CR}=2R_{\rm c} \chi^{1/2}/Mc_{\rm
s}$, with $M$ being the Mach number, $\chi$ being the ratio of cloud to
ambient density, and 
$c_{\rm s}$ being the sound speed.
During a time shorter than $\tau_{\rm CR}$,
the magnetic pressure is comparable
to the ram pressure, while afterwards ram pressure dominates.
For the MS clouds, $M\sim2$, and assuming $\chi=100$ from
Section~\ref{s:halo-density}, $\tau_{\rm CR} \sim 
10$ -- 30 Myr, depending on whether the near or far distance to the clouds is
assumed. As ${\rm a~few} \times \tau_{\rm CR}$ is a significant fraction of 
the MS age this suggests that during a large fraction
of a cloud's life-time magnetic pressure may play a significant role. Inclusion
of magnetic pressure in the calculations in Section~\ref{s:halo-density} may
somewhat reduce the required values for the Halo density.\\
 
(3) {\it Hydrodynamical approach}\\
Full hydrodynamical, 3-D simulations of HVCs moving through a diffuse
hot gaseous component were performed recently by Quilis \& Moore (2001).
They investigated both pure gas clouds, being in pressure
equilibrium with the external hot medium, and dark matter dominated
clouds with an additional potential field that maintains their dynamical
equilibrium. 
Particular cometary morphology, seen in the case of many HVCs as well as in
many MS clouds, was reproduced for both cloud types but under 
the condition that the diffuse medium has density $n_{\rm h}> 
10^{-4}$ cm$^{-3}$.

\subsection{Terminal velocity and neutral gas fraction}

At the tip of the MS, clouds have had plenty 
of time to decelerate by a drag force (Murali 2000).
The present velocity of the MS clouds is moderately supersonic 
(220 \kms~vs. sound
speed c$_s$ of $\sim 100$ \kms) suggesting that clouds
are most likely now at the terminal velocity (Benjamin and Danly 1997).
This means that clouds have stopped decelerating and that their
boundary layers are subject to Kelvin-Helmholtz instabilities, while
Rayleigh-Taylor instabilities are not important at this stage.
Benjamin \& Danly (1997) show that for a population of clouds with
known distances, terminal velocities and column densities, it is
possible to constrain the mean
density of the gaseous halo and cloud neutral
fraction $f_{\rm c}=N_{\rm HI}/(N_{\rm HI}+N_{\rm HII})$:
\begin{equation}
C_{\rm D}f_{\rm c}n_{\rm h}=\frac{2N_{\rm HI}g(z)}{v_{\rm T}^{2}},
\end{equation}
here $C_{\rm D}$ is the drag coefficient, $g(z)$ is the 
gravitational acceleration, and $v_{\rm T}$ is the cloud's 
terminal velocity. The mean column density for the MS
V and MS VI clouds is $\langle \log N_{\rm HI} \rangle \approx 19 $
cm$^{-3}$. Assuming that $v_{\rm T}=220$ \kms, the right-hand side of the
equation does not depend on distance and has a mean value of $\langle
C_{\rm D}f_{\rm c}\rho_{\rm h} \rangle \approx 6\times
10^{-4}$ cm$^{-3}$.
If we assume $C_{\rm D}=1$ and apply Halo densities from
Section~\ref{s:halo-density}, an estimate of the cloud's neutral fraction
can be obtained: at $z=15$ kpc $f_{\rm c}=0.6$, while at $z=45$ kpc $f_{\rm
c} \geq 1$.

\subsection{An alternative, gravitational confinement}
\label{s:alternative-interpretation}

An alternative explanation for cloud confinement is 
that they are gravitationally bound. This paradigm has been 
proposed for several compact HVCs (CHVCs)
which are in some respects similar to MS V and MS VI clouds. An additional
hint for this is provided by the peculiar velocity field seen in the MS VI
region (Figs. 3 and 4), and discussed in Section 4.1, that may be interpreted as
being due to differential rotation.  In the case of three CHVCs, Braun
\& Burton (2000) found evidence for such orbital motion.  In order to compare
the MS VI cloud with these CHVCs we have attempted to model the
velocity field north of Dec 12\degree 30$'$ using the tilted ring
algorithm. The derived rotation curve has a quick and almost linear rise
to $\sim$ 15 \kms, and then flattens out with large uncertainties.

This curve was used to 
obtain an estimate of the dynamical mass:
$M_{\rm dyn}=1.5\times10^{7}$ M$_{\odot}$ or  $M_{\rm dyn}=5\times10^{6}$
M$_{\odot}$, depending whether the far or near distance is assumed.
This further implies dark-to-visible mass
ratios of 50 to 180, respectively, after including a contribution by helium
of 40\% by mass to the observed HI mass.  Therefore, if interpreted as a
rotating HI cloud, this would require a very dark matter halo.

No model for the formation of the MS has ever predicted the existence of
dark-matter dominated clouds in this gas filament.  Even
the parent galaxies, the MCs, are not rich in dark matter. It is
therefore hard to believe that rotational support is realistic.
This was the conclusion of Section~\ref{s:cloud-stability}.
However the derived rotation curve is 
very similar to the rotation curves of
CHVCs derived by Braun \& Burton (2000).  
Dynamical masses of these CHVCs are about 10 times higher than the
dynamical mass of the MS VI region, while the dark-to-visible mass ratio of
the MS VI region is a few times higher than that for CHVCs.  As the
velocity field of MS VI is most likely the result of  clump blending along
the line-of-sight, it is interesting to note how  well this effect can mimic
differential rotation.

\subsection{On the clump morphology}
\label{s:clump-morphology}

\begin{figure*}
\epsscale{1.0}
\plotone{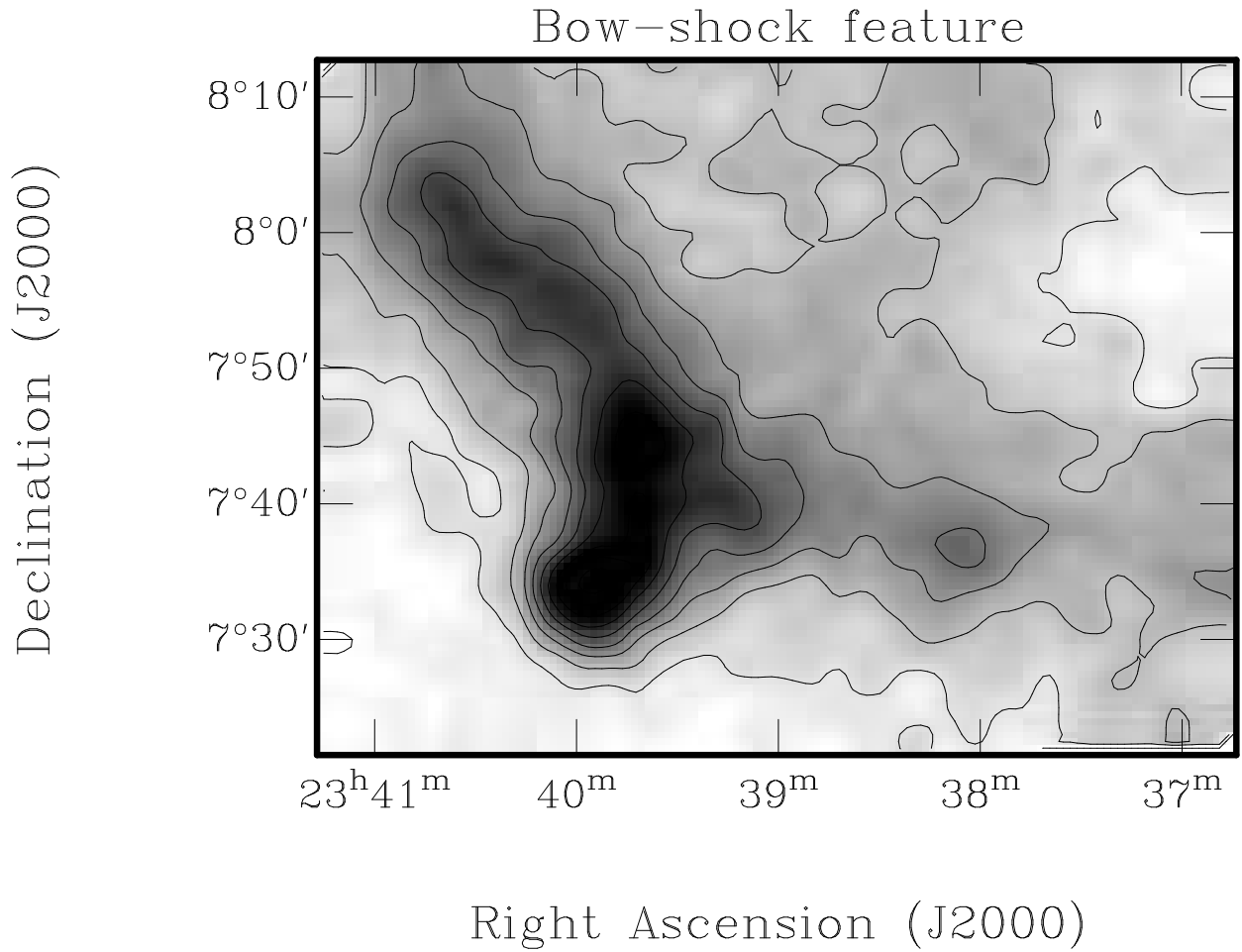}
\caption{\label{f:feature} An HI intensity image of the bow-shock-like 
feature in the MS V region at the LSR velocity of $-351$ \kms.}
\end{figure*}
 
Several clumps in the MS data cube show gradients in
the brightness and velocity distribution, exhibiting a cometary or
head-tail morphology. The best example is the
south-eastern side of the MS V cloud
at $-340$  to $-355$ \kms, see Fig.~\ref{f:feature}. 
This bow-shock-like feature may be a strong compression front,
centered at RA 23$^{\rm h}$ 40$^{\rm m}$, Dec 07\degree 34$'$ 
with two tails swept back on either side.
This is strongly suggestive of a supersonic 
interaction between the HI cloud and an external, low density, ionized
medium.  

Similar cometary morphology is seen in many HVCs 
in surveys of \cite{Meyerdierks91} and \cite{Bruns00}. 
\cite{Putman-thesis} found many such clouds in the MS, which she
speculated may be in the process of evaporation. 
The 3-D hydrodynamical simulations of an HVC moving
through a diffuse hot medium by Quilis \& Moore (2001) show a 
bow shock in front of the cloud, gas
compression at the front edge of the cloud, and visible or invisible tails
behind the cloud, depending on the density of the ambient medium. 
The double-tail morphology seen here is more similar to 
the simulations of the interaction between a disk-like structure 
with the surrounding medium by \cite{Quilis00}. 
These simulations were the first successful attempt to incorporate complex
turbulent and viscous stripping at the interface of the cold and hot
gaseous components. They predict a compression front and
two tails folding on both sides. 
The similarity of the bow-shock-like feature with the simulations by 
\cite{Quilis00} demonstrates the importance of turbulent and viscous 
mixing at the boundary layers when considering
cloud -- ambient medium interactions and should be fully explored in the
future.

\section{Conclusions}
\label{s:conclusions}
In this paper, we have presented new HI observations of two regions, MS V
and MS VI, at the northern tip of the MS using the Arecibo telescope. 
The new data sets show the complex morphology of the MS with numerous
inter-connected clumps. 
We have found double velocity structures in the MS V region. This region
also contains an interesting bow-shock-like feature that is strongly
suggestive of an interaction between the MS and an external medium.
The MS VI data set shows a large velocity gradient, caused most
likely by blending of small clumps along the line of sight. This region
resembles the velocity field of a rotating disk, but we 
find that rotation is not a likely explanation.

Several MS clumps have been isolated to investigate their 
confinement mechanism. 
We show that unreasonably large amounts of dark matter are 
required in order for clumps to be gravitationally confined. 
Clumps do not seem to be in free expansion either. The easiest
way to explain the clump properties is with external pressure
confinement by the hot Galactic Halo. In this scenario we place an upper limit
on the Halo density: $n_{\rm h} = 10^{-3}$ cm$^{-3}$ at $z \approx 15$ kpc,
and $n_{\rm h} = 3 \times 10^{-4}$ cm$^{-3}$ at $z \approx 45$ kpc.
These values agree well with several previous indirect 
observations of the Halo, but are significantly higher than the 
values predicted theoretically for an isothermal hot Halo by Wolfire et
al. (1995). Our results are closer to the expected Halo densities of the
Moore \& Davis model. 

Although cloud evaporation is most likely an important process
at the tip of the MS,
the mass-loaded approach for cloud interaction with the ambient medium
requires very high mass-loss rates to enable the existence of 
the hot Halo gas. The ballistic consideration of clouds in the Halo 
suggests that during a large fraction of the cloud lifetime
magnetic pressure may have a significant role in cloud
evolution. Inclusion of magnetic pressure would reduce somewhat 
the required values for the Halo density. 

Hydrodynamical modeling of cloud interaction with the ambient
medium also requires high Halo densities to reproduce cometary features
commonly seen in HI observations of HVCs. If we assume that MS clouds are 
at their terminal velocity, the
cloud's neutral fraction can be estimated at $z=15$ kpc $f_{\rm c}=0.6$, 
while at $z=45$ kpc $f_{\rm c} \geq 1$.

\acknowledgements{}
We wish to thank several colleagues for their valuable input 
and discussions. We
are in debt to Phil Perillat and Chris Salter for helpful suggestions 
during observations and data reduction.
We are grateful to Abby Hedden for eagerly helping with observations and
to Amanda Kirchner for assisting during the June 2000 observing run. We
thank Mary Putman for stimulating discussions and for providing the HIPASS
MS data at the northern tip prior to publication. 
Discussions with Tom Jones, Bon-Chul Koo, Robert Braun and Butler Burton are
greatly appreciated. We also thank Chris Salter and Lister Staveley-Smith 
for a careful reading of the paper which resulted in many
improvements.

\label{lastpage}
\end{document}